\documentclass[submit]{epsv8_submissiondatespecification}
\usepackage[pdftex]{graphicx}

\usepackage{lineno}
\title{A comparative study on the self-similarity hypothesis of regular and slow earthquake growth}
\author{
Dye SK Sato, Research Institute for Marine Geodynamics, Japan Agency for Marine-Earth Science and Technology,
3173-25 Showa-machi, Kanazawa-ku
Yokohama Kanagawa 236-0001, Japan,
daisukes@jamstec.go.jp
}

\abstract{
Fault ruptures of regular earthquakes typically grow in a self-similar manner, where the radiated energy is proportional to the seismic moment. Their proportionality factor, termed as scaled energy, has been conventionally described as the ratio of stress drop to stiffness. By analyzing the self-similar circular crack model by Sato and Hirasawa (1973), Matsu'ura (2024) found a correction prefactor for this theoretical representation of the scaled energy, the cubed ratio of the rupture speed to the S-wave speed. The stress drop times the cubed rupture speed is the scaling prefactor of the self-similar seismic moment, and thus, Matsu'ura's solution tells that the seismic moment rate is determined by the scaled energy in the self-similar rupture growth. We rearrange the properties of the self-similar solution from this perspective, apply this Sato-Hirasawa-Matsu'ura relation to a series of seismic events thought to be self-similar, and estimate their source parameters and scaled energies. According to our estimation using the above self-similarity model, seismologically detected low-frequency and very-low-frequency earthquakes have a common scaled energy, while geodetically detectable slow slip events indicate multiple modes with different scaled energies. Meanwhile, for very-low-frequency earthquakes and tectonic tremors, the seismic moments were significantly smaller than expected for self-similar ruptures despite their moments being proportional to cubed duration values. It keeps alive a hitherto less contemplated possibility: the proportionality of the moment to cubed duration may not be a manifestation of the self-similarity for slow earthquake rupture growth.
}


\begin{document}

\maketitle

\section{Preliminaries}
The definition of the scale of earthquake rupture is founded on two quantities with energy dimension, tied to waves and their radiated sources, slips. 
One is the kinetic energy released by the seismic motion, known as the radiated (seismic) energy $E_{\rm R}$:
\begin{equation}
    E_{\rm R} =\frac 1 2 \int dV \rho \dot u^2,
    \label{eq:radiatedenergy}
\end{equation}
where $\rho$ and $u$ denote the mass density and displacement of the medium, respectively; 
\citet{haskell1964total} convention dropping the $1/2$ prefactor of the radiated energy (eq.~\ref{eq:radiatedenergy}) is often adopted in practice~\citep{matsu2024reconsideration}, but here it is corrected.
In general, the seismic source is the eigenstress, which is the product of inelastic strain (eigenstrain) multiplied by stiffness~\citep{backus1976momentI,backus1976momentII}.
Far from the fault rupture concentrating inelastic strains on and around a fault plane, eigenstress approximates the seismic moment $M_0$, which is the stiffness times the total amount of slip $\delta$ released at the rupture~\citep{matsu2019physical}.
For shear slip with a spatially uniform direction, 
\begin{equation}
    M_0= \mu \int d\Sigma \delta,
\end{equation}
where $\mu$ is the rigidity of the medium enough away that the fracture zone can be regarded as a crack face. 
Radiated energy and seismic moment are both observables measurable from waves, representing essential features of dynamic seismic motion and static faulting deformation.

In the celebrated course of relating the energy of quakes $E_{\rm R}$ to the size of sources $M_{\rm 0}$ to arrive at the earthquake moment magnitude~\citep{hanks1979moment}, 
\citet{kanamori1975theoretical} found proportionality between measured $E_{\rm R}$ and $M_0$; from recent analysis of \citet{kanamori2020estimation}, with accounting for the \citet{haskell1964total} convention,
\begin{equation}
\frac {E_{\rm R}} {M_0}\approx 0.2\times 10^{-4}  
\hspace{10pt}({\rm Regular\,EQs}).
\label{ea:scaledenergyEQs}
\end{equation}
This $E_{\rm R}/M_0$, called scaled energy~\citep{kanamori2006energy}, is expressed in terms of physical quantities on the fault plane (source parameters) by assuming specific mechanical processes.
The simplest model will be a circular crack with a uniform stress drop $\Delta \tau$ expanding at a uniform constant rupture speed $v_{\rm r}$ close to the S-wave speed $v_{\rm S}$, where 
the scaled energy is~\citep{kanamori1975theoretical,kanamori1977energy}
\begin{equation}
    \frac {E_{\rm R}} {M_0}\simeq \frac 1 2 \frac{\Delta \tau}{\mu} \hspace{10pt}({\rm Self{\mbox -}similar\,cracks},\, v_{\rm r}\sim v_{\rm S}).
    \label{eq:Tsuboi}
\end{equation}
Thus, it is recognized that, as long as a nearly isotropic self-similar rupture is considered a faulting process, the stress drop of the fault plane is approximately ten-thousandth of the stiffness, an almost constant value around a few MPa~\citep{kanamori1975theoretical,allmann2009global}.

Since then, seismology has linked to the working hypothesis that equates the final sizes of earthquakes to the sizes of one self-similarly propagating crack at various periods---regarding various earthquakes as terminations of an almost common self-similar rupture.
Recent analyses show that this hypothesis of self-similarity breaks down for large earthquakes but is valid up to around M${}_{\rm W}$ 6--7~\citep{uchide2010scaling,meier2016evidence,denolle2016new,meier2017hidden}.
Here, saying that a solution is self-similar means that the equation and the boundary shape and condition for that solution are invariant in a scale transformation. Since length and time scales are proportional in the elastodynamic equation of motion, the seismic moment, proportional to cubed distance, is proportional to the cube of duration $t$ if self-similar.
According to \citet{uchide2010scaling},
\begin{equation}
    M_0(t)/t^3 \approx 2\times 10^{17}\,\,[{\rm Nm/sec}^3] \hspace{10pt}({\rm Regular\,EQs}).
\end{equation}
For a self-similar circular crack with constant rupture speed~\citep{sato1973body}, 
\begin{equation}
    M_0(t)\simeq \frac{16}{7}\Delta \tau (v_{\rm r}t)^3\hspace{10pt}({\rm Self\mbox{-}similar\,cracks}).
    \label{eq:SatoHirasawa}
\end{equation}
The problem setting of \citet{sato1973body} is an approximate application of Eshelby's (1957) elastostatic solution to elastodynamic problems, exploiting the essence of fault rupture, thus known as the Sato-Hirasawa model.
The model calculation by \citet{sato1973body} 
mostly explains the observation of \citet{uchide2010scaling}, when using the above-mentioned $\Delta\tau\sim$ MPa as well as 3.5 km/sec often assumed for $v_{\rm r}$. 
This agreement between reality and the simplified model will be striking, given the order-level uncertainty of stress drop estimates~\citep{allmann2009global}.

Note that self-similarity does not hold for rupture termination/arrest. Moreover, even after rupture propagation stops, non-self-similar fault slip continues for a certain period, releasing  moments of the same order as during rupture propagation~\citep{madariaga1976dynamics,meier2017hidden}. The complexity of rupture arrest is outside the scope here, but we proceed on the understanding that we are discussing order-level accuracy with neglecting the contribution of non-self-similar slips of rupture arrest contained in the net moment.

As such, the self-similarity hypothesis has been verified in fast fault ruptures. 
Interestingly, though not initially claimed so, self-similar scaling curves have been detected also in slow earthquakes~\citep{ide2008bridging,maeda2009spatiotemporal,gomberg2016reconsidering,ide2018seismic,michel2019similar,frank2019daily,supino2020self,takemura2024revisiting}. 
The earliest study may be \citet{ide2008bridging}, 
which reported that the scaled energy of very-low-frequency earthquakes and tectonic tremors was constant and about five digits smaller than that of regular earthquakes. Although the constancy of the scaled energy was not a sufficient condition for the rupture self-similarity, the resemblance between regular and slow earthquakes in terms of energetics was already anticipated here.
Subsequently, a series of slow earthquakes were found to indicate the proportionality of moments to the cubed duration, that is, apparent self-similarity~\citep{gomberg2016reconsidering,michel2019similar,frank2019daily,supino2020self}. 

Nonetheless, when considering the moment scaling for the slow earthquake family, the seismic moment is roughly proportional to the squared duration, namely, diffusive~\citep{ide2007scaling}. 
This Ide's scaling for the slow earthquake family has long been subject to debate~\citep{peng2010integrated,ide2023slow}, 
but it is almost consensual that the unified relationship between the moment and duration of the slow earthquake family is non-self-similar. 
The self-similar moment series of respective slow earthquakes are therefore not situated on a common scaling curve, and the proportionality factor for moment and cubed duration is specific to each slow earthquake activity. In this sense, the apparent self-similarity of slow earthquakes is dissimilar to the ``genuine'' self-similarity of regular earthquakes. 

Various macro-/micro-physical models for the diffusive scaling have been presented ~\citep[e.g.,][]{ando2010slip,ben2012episodic,yamashita2013generation} since \citet{ide2008bridging} recognized the diffusivity of the moment-duration scaling in \citet{ide2007scaling}. 
Meanwhile, concerning the apparent self-similarity of dynamic rupture growth with different velocities commonly observed in individual slow earthquakes, whereas attention has been drawn to the possible scaling crossover from self-similar one to diffusive one as in regular earthquakes~\citep{gomberg2016reconsidering,ide2023slow}, 
there is not much consideration on the apparently self-similar growth itself.

For the slow earthquakes, we cannot rule out the possibility that the self-similar scaling is pseudo~\citep{ide2023slow}. 
That is why, paradoxically, we now dare to assume that various regular and slow earthquakes are hypothetically self-similar, to consider the conditions for those earthquakes to be self-similar. In other words, this study considers how to make the self-similarity of slow rupture growth a verifiable/refutable proposition as in regular earthquakes. 

Hereafter, we shall restrict our discussion to this self-similar rupture growth.
The remaining question is then how to understand the proportionality factor of the moment and cubed duration that varies from phenomenon to phenomenon.
The formula (eq.~\ref{eq:Tsuboi}) of \citet{kanamori1977energy} is unfortunately not applicable to $v_{\rm r}/v_{\rm S}\ll1$, as can be inferred from the fact that $E_{\rm R}$ must be zero for the quasi-static limit of $v_{\rm r}/v_{\rm S}\to0$, and there have been no simple measures for comparing faulting phenomena with far different rupture speeds. There, \citet{matsuura2024A} generalized eq.~(\ref{eq:Tsuboi}) by solving the self-similar crack problem of \citet{sato1973body} with unrestricted $v_{\rm r}/v_{\rm S}$ values: 
\begin{equation}
    \frac {E_{\rm R}} {M_0}\simeq \frac 1 2 \frac{\Delta \tau}{\mu}\left(\frac{v_{\rm r}}{v_{\rm S}}\right)^3 \hspace{10pt}({\rm Self\mbox{-}similar\,cracks}),
    \label{eq:Matsuura}
\end{equation}
This concise expression states that radiation efficiency~\citep{savage1971relation,sato1973body} is the cubed ratio of the rupture speed to the S-wave speed for the self-similar rupture growth. 
\citet{supino2020self} had reached a similar $v_{\rm r}/v_{\rm S}$-dependence from the numerical solution of the \citet{sato1973body} model, which is now given a solid basis by \citet{matsuura2024A}.

One may notice an unexpected relation that the theoretical equations for dynamic scaling of moments (eq.~\ref{eq:SatoHirasawa}) and scaled energy (eq.~\ref{eq:Matsuura}) are commonly parametrized by the product $\Delta \tau v_{\rm r}^3$ of the stress drop and cubed rupture speed. 
The other coefficients ($\mu$, $v_{\rm S}$, and crack-shape-dependent prefactors) are approximately constant for many seismic events, with little involvement in rupture processes.
Once this is realized, the theoretical solution (eq.~\ref{eq:Matsuura}) of \citet{matsuura2024A} for the Sato-Hirasawa model suggests a nontrivial relationship between the dynamic moment scaling and scaled energy (eq.~\ref{eq:SatoHirasawaMatsuura} below). 
By using it, as for regular earthquakes, 
we can discuss the self-similarity for slow earthquakes from the scaling prefactors, not relying solely on scaling exponents (cubed, squared, and so on). We applied this Sato-Hirasawa-Matsu'ura relation (eq.~\ref{eq:SatoHirasawaMatsuura}) to actual data and investigated the source parameters and scaled energies for various apparent self-similar ruptures with different velocities to examine their apparent self-similarity to be tested.

\section{Sato-Hirasawa-Matsu'ura Relation and Kinematic Implications}

It is a virtue of physics-based models that they relate observed data to physical properties of faults, but for this reason, their predictions are often not closed only with measurable quantities.
For example, in eqs.~(\ref{eq:SatoHirasawa}) and (\ref{eq:Matsuura}), $\Delta\tau$ and $v_{\rm r}^3$ appear in the form  of a product, unable to be estimated independently of each other. 
Therefore, we now eliminate the source parameters from these equations to obtain a relationship between observables free from assumptions on the rupture processes. From eqs.~(\ref{eq:SatoHirasawa}) and (\ref{eq:Matsuura}),
\begin{equation}
    \frac{E_{\rm R}}{M_0}=\frac{7}{32}\frac{M_0}{\mu(v_{\rm S}t)^3}
    \hspace{10pt} ({\rm Sato{\mbox-}Hirasawa{\mbox-}Matsu'ura}).
    \label{eq:SatoHirasawaMatsuura}
\end{equation}
The coefficients $\mu$ and $v_{\rm S}$ are almost constant in seismogenic zones; $\mu$ is around 30 GPa, and $v_{\rm S}$ is around 4 km/sec. 
Thus, eq.~(\ref{eq:SatoHirasawaMatsuura}) represents an almost closed relation between observables. 
Equation(\ref{eq:SatoHirasawaMatsuura}) predicts that, surprisingly,
the relationship between the scaled energy and 
the dynamic scaling of seismic moments is determined by the properties of the medium outside the fault plane, independent of the fault properties. 
Let this nontrivial theoretical prediction be called the Sato-Hirasawa-Matsu'ura relation.

Equation (\ref{eq:SatoHirasawaMatsuura}) is intelligible in essence when viewed in relative values for different rupture speeds $v_{\rm r}$ and $v_{\rm r}^\prime$:
\begin{equation}
    \frac{(E_{\rm R}/M_0)(v_{\rm r})}{(E_{\rm R}/M_0)(v_{\rm r}^\prime)}=\frac{M_0(t;v_{\rm r})}{M_0(t;v_{\rm r}^\prime)},
    \label{eq:SatoHirasawaMatsuura_relative}
\end{equation}
where the dependence on rupture speed $v_{\rm r}$ and duration $t$ is specified explicitly; duration $t$ here refers to the time elapsed from the onset from which the self-similar scaling well fits the data in practice, where the true rupture initiation time is unknown.  
The scaled energy $E_{\rm R}/M_0$ is constant in time for self-similar cracks, and the ratios of time derivatives of its numerators and denominators are also:  
\begin{equation}
    (\dot E_{\rm R}/\dot M_0)(v_{\rm r}) =(E_{\rm R}/ M_0)(v_{\rm r}).
\end{equation}
Similar relationships hold for higher-order derivatives.
$E_{\rm R}/M_0$ is often substituted with $\dot E_{\rm R}/\dot M_0$ in slow earthquake analysis.

The meaning of eq.~(\ref{eq:SatoHirasawaMatsuura_relative}) becomes even clearer when eq.~(\ref{eq:SatoHirasawaMatsuura_relative}) is read as a relation between $E_{\rm R}$ and $M_0$: 
\begin{equation}
    \frac{M_0(t;v_{\rm r})}{M_0(t;v_{\rm r}^\prime)}=\sqrt{\frac{E_{\rm R}(t;v_{\rm r})}{E_{\rm R}(t;v_{\rm r}^\prime)}}.
    \label{eq:moment_radiation_ratio_relation}
\end{equation}
The dependence of $E_{\rm R}$ and $M_0$ on duration and rupture speed is now explicitly written. 
Equation (\ref{eq:moment_radiation_ratio_relation}) shows that, given a duration value, relative differences are equal between seismic moments the square roots of radiated energies. 
Because seismic moments are proportional to fault slips, eq.~(\ref{eq:moment_radiation_ratio_relation}) is a macromechanical prediction to
relate magnitude differences in fault slips to magnitude differences in seismic waves.

To conclude this section, we consider the theoretical generality and practical applicability of the model on which the Sato-Hirasawa-Matsu'ura relation is based. 
The model now assumes constant stress drop with spatial uniformity, but 
this is the unique choice for the deterministic self-similar models excluding stochastic effects~\citep{hirano2022source} so that 
the Sato-Hirasawa model may be regarded as the typical solution of self-similarity. 
Rupture propagation directivity is here neglected, but the same seems to hold for ellipses using an isoarea mapping~\citep[i.e., reading products of major and minor radii of ellipses as cubed radii of perfect circles;][]{kaneko2015variability}. 
Fracture energy only plays a role in determining $v_{\rm r}$ in classical fracture mechanics. 
The effect of dissipation (heat) also has no room to appear in waves in self-similar problems, where absolute stress differences are eliminated from the elastodynamic equation of motion and boundary conditions. 
Concerning the order-level applicability to actual data, 
we now neglect the effect of rupture arrest of the same-order as mentioned in preliminaries. 
Rupture initiation locations and rupture zone shapes do not change results by one order of magnitude unless extreme cases are considered~\citep{kaneko2015variability}. 
The applicability limits of the approximate use  of elastostatic solutions by \citet{eshelby1957determination} in elastodynamic problems are discussed in the last section.

\section{Applying Theoretical Relations to Earthquake Growth with Different Rupture Speeds}

Examples of slow earthquakes include low-frequency earthquakes (LFEs), tectonic tremors, very-low-frequency earthquakes (VLFEs), and slow slip events (SSEs). In this section, we list their clusters reported to indicate apparent self-similar scalings, to investigate their scaled energy, rupture speed, and stress drop, as well as the self-similarity hypothesis of each member of the slow seismic family. The outline of the slow earthquake family in this section owes much to \citet{nishikawa2023review}.

The model application procedure is as follows.
The self-similar crack model predicts that the moment-duration plot will be placed on a single scaling curve if corrected for $\Delta\tau  v_{\rm r}^3$ or scaled energy; 
from eqs.~(\ref{eq:SatoHirasawa}) and (\ref{eq:SatoHirasawaMatsuura}), 
\begin{equation}
    t^3=\frac{7}{16\Delta \tau v_{\rm r}^3}
    M_0(t)=\left.\frac{7}{32\mu v_{\rm S}^3}M_0(t)\right/\frac{E_{\rm R}}{M_0}.
    \label{eq:scaledmoment}
\end{equation}
For Regular earthquakes, scaled energy $E_{\rm R}/M_0$ and $\Delta\tau v_{\rm r}^3$ (the $M_0/t^3$ prefactor)
have been measured and confirmed to take consistent values. 
For LFEs and SSEs, $\Delta\tau v_{\rm r}^3$ has been measured ~\citep{michel2019similar,supino2020self,tan2020connecting}, and then we can estimate their scaled energies using Matsu'ura's solution (eq.~\ref{eq:Matsuura}). 
For VLFEs, the scaled energy has been measured~\citep{ide2018seismic}, and then we can estimate their $\Delta\tau v_{\rm r}^3$ using Matsu'ura's solution (eq.~\ref{eq:Matsuura}).

Attention should be drawn to the fact that self-similarity is still a hypothesis for slow earthquakes. 
If the scaled energy is independently measured, 
this hypothesis can be tested by checking whether the $M_0/t^3$ curve normalized by the scaled energy $E_{\rm R}/M_0$ is on the scaling curve (eq.~\ref{eq:scaledmoment}) of the self-similar mechanics, even though it is a rough argument based on a simple model.
The VLFE fits this situation and hence can be used to verify the self-similarity hypothesis. 
Conversely, if $\Delta\tau v_{\rm r}^3$ is measured, the self-similarity hypothesis can be verified by checking whether the predicted scaled energy converted from $\Delta\tau v_{\rm r}^3$ matches the measured one. 
Scaled energy has not been measured in LFEs as far as we know, while \citet{maeda2009spatiotemporal} measured it for SSEs, on which we can discuss the self-similarity hypothesis.

Therefore, we firstly look at the source parameters and scaled energy of each member of the slow earthquake family based on the working hypothesis of self-similarity as in previous studies. Secondly, we reexamine this conversion results, by using different data, and assess the self-similarity hypothesis on slow earthquakes. 

Figure~1 shows a moment-duration plot rescaled by eq.~(\ref{eq:scaledmoment}). 
Relative positions before rescaling are shown in grey for comparison using regular earthquake positions rescaled. 
In data conversion, typical values of stiffness of 30 GPa, P-wave speed of 7 km/sec, and S-wave speed of about 4 km/sec are assumed (at this time, the regular earthquake rupture speed is about 3.5 km/sec). Similar values were used in the literature from which we took data. 
The moments of the slow earthquake family, where the final moments were approximately bounded by the diffusive scaling (pink in Fig.~1), merge approximately into a single self-similar scaling curve after the rescaling of eq.~(\ref{eq:scaledmoment}).
Numbers in parentheses in Fig.~1 represent nominal estimates of the scaled energy for respective clusters calculated below.

First, we look at $\Delta\tau  v_{\rm r}^3$ and scaled energy estimates under the self-similarity hypothesis. 
The data of regular earthquakes are borrowed from the compilation of \citet{allmann2009global}, 
containing their own analysis and 
\citet{hough1996observational}, \citet{boatwright1994regional}, \citet{mori1990source}, 
\citet{tajima2007seismic}, 
\citet{humphrey1994seismic}, 
\citet{archuleta1982source}, 
\citet{venkataraman2004observational}, and \citet{abercrombie1995earthquake}. 
The data of \citet{allmann2009global} is in the frequency domain, 
and now they are converted to the time domain 
so that the center of the self-similar scaling band of \citet{allmann2009global} becomes consistent with the prediction of the self-similar model (eq.~\ref{eq:SatoHirasawa}). In short, we now fitted $\Delta\tau  v_{\rm r}^3$ of regular earthquakes. 
Conventional analyses of regular earthquakes convert the corner frequencies to $v_{\rm r}t$ using the model of \citet{madariaga1976dynamics}, and then $\Delta\tau $ are fitted from the $M_0/t^3$ curve, 
but we end up doing a similar fit of $\Delta \tau v_{\rm r}^3$. Regular earthquakes are known to be generally self-similar as mentioned in preliminaries.

Tectonic tremors and VLFEs can often be interpreted as the same events detected in different frequency bands~\citep{ide2007scaling,kaneko2018slow,masuda2020bridging}. 
At this time, given the difference in the measurement frequency band between VLFEs (often 0.02--0.05 Hz) and tectonic tremors (often 1--8 Hz), 
moments of given events are approximated by that of VLFEs of lower frequencies, while their radiated energies are approximated by that of tectonic tremors of higher frequencies. 
Under such an understanding, \citet{ide2008bridging} measured scaled energies of VLFEs and tectonic tremors and found their scaled energy constant; according to \citet{ide2014universality},
\begin{equation}
    \frac{E_{\rm R}}{M_0}\approx 0.3\times 10^{-9}\hspace{10pt}({\rm VLFEs\,\&\,Tremors}).
    \label{eq:scaledenergyVLFEs}
\end{equation}
This is approximately $10^{-5}$ times the scaled energy of an earthquake. 
Figure~1 borrows relatively larger datasets~\citep{ide2008bridging,matsuzawa2009source,ide2014universality,ide2015thrust,maury2016comparative} included in the compilation (stacked: yellow; unstacked: blue) of \citet{ide2018seismic}.
When cite effects are corrected, this value is depth-independent within a moment rate range from $10^{11}$ to $10^{15}$ [Nm/sec]~\citep{takemura2024revisiting}.

Strictly speaking, \citet{ide2014universality} reported that the $2E_{\rm R}/M_0$ estimate of VLFEs and tectonic tremors is $0.5\times 10^{-9}$. 
The \citet{haskell1964total} convention (mentioned below eq.~\ref{eq:radiatedenergy}) is inherited to the widely-used scaled energy expression $\int dt \ddot M_0^2/(10\pi\rho v_{\rm S}^5)$ from \citet{rudnicki1981energy}. See \citet{matsu2024reconsideration} for details.

Assuming self-similarity, we can estimate the corresponding source parameters for VLFEs and tectonic tremors as $\Delta \tau v_{\rm r}^3 \simeq (1/2) \mu v_{\rm S}^3 \times 10^{-9}$ from eq.~(\ref{eq:scaledenergyVLFEs}) via eq.~(\ref{eq:Matsuura}). 
The associated stress drop is 
about 6 MPa for the rupture speed of rapid tremor reversals (200 km/h); the stress drop is unrealistically higher even than rigidity for rupture speeds close to the migration speed (10 km/day). 
Given that the stress drop of a VLFE is unlikely to be higher than that of a regular earthquake, 
one likely picture will be that a fast rupture like rapid tremor reversals propagate over a single patch while the migration is a rupture cascade~\citep{ellsworth1995seismic} of multiple patches.

Concerning LFEs, \citet{supino2020self} estimated the possible values of source parameters by assuming the LFEs of $M_0\propto t^3$ they investigated followed the Sato-Hirasawa model (eq.~\ref{eq:SatoHirasawa}) as
\begin{equation}
    \frac{\Delta \tau}{\mu}\left(\frac{v_{\rm r}}{v_{\rm S}}\right)^3\approx 1\times 10^{-9} \hspace{10pt}({\rm Self{\mbox-}similar\,LFEs}).
    \label{eq:2scaledenergyLFEs}
\end{equation}
Figure~1 converts the corner frequency and duration so that the data of \citet{supino2020self} satisfy 
Sato-Hirasawa's moment dynamic scaling (eq.~\ref{eq:SatoHirasawa}) when using $(\Delta\tau/\mu)(v_{\rm r}/v_{\rm S})^3$ values \citet{supino2020self} obtained, that is, by using the same conversion \citet{supino2020self} used. 

Given eq.~(\ref{eq:Matsuura}), this  $(\Delta\tau/\mu)(v_{\rm r}/v_{\rm S})^3$ value (eq.~\ref{eq:2scaledenergyLFEs})
suggests that the scaled energy of LFEs is almost the same as that of VLFEs and tectonic tremors (eq.~\ref{eq:scaledenergyVLFEs}), as long as the self-similarity is assumed for LFEs. 
This is a natural value, given that LFEs are considered to be the building blocks of tremors or VLFEs~\citep{shelly2007non,gomberg2016reconsidering}.
Stress drop and rupture speed will also take similar values for LFEs and VLFEs.

For SSEs, \citet{michel2019similar} estimated the following for an event of $M_0\propto t^3$ (cyan in Fig.~1, unstacked) they scrutinized with geodetic data: 
\begin{equation}
\begin{array}{c}
    \Delta\tau\approx 6[{\rm kPa}]
    \\
    v_{\rm r}\approx 4[{\rm cm/sec}]
\end{array}
\hspace{10pt}({\rm Self{\mbox-}similar\,SSEs\,1}).
\end{equation}
Precisely, \citet{michel2019similar} unmentioned the rupture speed associated with circular crack fitting, and we estimated it from their moment-area and moment-duration plots.
From the above source parameters,
\begin{equation}
    \frac{\Delta \tau}{\mu}\left(\frac{v_{\rm r}}{v_{\rm S}}\right)^3\simeq 2\times 10^{-22}. \hspace{10pt}({\rm Self{\mbox-}similar\,SSEs\,1}).
\end{equation}
According to Matsu'ura's solution (eq.~\ref{eq:Matsuura}),
the scaled energy of this self-similar SSE is around $1\times10^{-22}$, which is 17--18 digits smaller than that of regular earthquakes. 

\citet{tan2020connecting} estimate the source parameters of SSEs from LFE data, assuming the presence of an SSE behind a cluster of LFEs as in \citet{frank2019daily}, instead of directly measuring SSEs geodetically. 
\citet{tan2020connecting} reported two other self-similar SSE events, where moment-duration prefactors are different by two orders of magnitude (purple and orange in Fig.~1, stacked). 
Source-parameter estimates for respective clusters are~\citep{tan2020connecting}
\begin{equation}
\begin{array}{c}
    \Delta\tau\approx 6[{\rm kPa}]
    \\
    v_{\rm r}\approx 6[{\rm km/day}]
\end{array}
\hspace{10pt}({\rm Self{\mbox-}similar\,SSEs\,2})
\end{equation}
and
\begin{equation}
\begin{array}{c}
    \Delta\tau\approx 30[{\rm kPa}]
    \\
    v_{\rm r}\approx 708[{\rm km/day}]
\end{array}
\hspace{10pt}({\rm Self{\mbox-}similar\,SSEs\,3}).
\end{equation}
Precisely, \citet{tan2020connecting} obtained a median 6 kPa of stress drop for the larger cluster, but their data~\citep[Fig.~3D in][]{tan2020connecting} shows a systematic moment-dependence of stress drops so that we roughly estimated the stress drop for the smaller cluster from their figure. 
These estimated rupture speeds, 6 km/day and 40 km/hour, are the same as or slightly slower than those for usual SSEs and rapid tremor reversals, respectively. 
Interestingly, the longer-duration cluster of \citet{tan2020connecting} is similar to that of \citet{michel2019similar} in terms of both duration-dependent moments and source parameters. 

There is variability in the $M_0\propto t^3$ prefactors of SSEs, which suggests multiple modes of SSEs with different scaled energies. 
Under the self-similarity hypothesis, the scaled energies of SSEs corresponding to the above data are estimated to be $0.5\times 10^{-21}$ and $0.4\times 10^{-14}$, respectively, from eq.~(\ref{eq:Matsuura}). 
Some episodic tremors and slips also appear to exhibit $M_0\propto t^3$ scaling~\citep{nakamoto2021cascading}. 

Now, let us examine the estimation results obtained under the hypothesis of self-similarity, by using other clues to check this hypothesis.

Concerning the LFEs, the hypothesis of self-similarity returned a scaled energy estimate consistent with the observed scaled energy of VLFEs thought to be a cluster of LFEs. 
Then, the self-similarity hypothesis seems supported for LFEs.

Concerning the VLFEs and tectonic tremors,
their moments rescaled by scaled energy are smaller compared to the self-similar scaling. 
These systematic deviations are twice the variability of regular earthquakes, thus significant. 
It indicates that the data of VLFEs and tectonic tremors violate the prediction of self-similarity.
The trend of violation implies that the observed scaled energy in these activities is too larger than the self-similarity expectation: apparently, the VLFEs emit relatively too strong waves to be called self-similar.
There are three possible interpretations of this inconsistency. Two of them defend the self-similarity hypothesis. 
The first is to regard deviations from predictions of self-similarity as estimation errors (moment underestimation or duration overestimation).
The second one seeks reasons for the rupture arrest after self-similar growth.
The last interpretation is the concerned possibility that self-similarity is only apparent for slow earthquakes even though they exhibit the scaling (exponent) of $M_0\propto t^3$~\citep{ide2023slow}.
Although the pursuit of its cause is beyond the scope of this study, 
the self-similarity violation of VLFEs and tectonic tremors is one of the interesting results we found in this study.

Concerning the SSEs, 
\citet{maeda2009spatiotemporal} reported that when the radiated energies of SSEs are evaluated from synchronized tectonic tremors, the scaled energy of SSEs is as high as VLFEs' in their observation. 
This is far larger than the scaled energy estimated here by assuming the self-similarity and implies the same sense as VLFEs: too much wave radiation.
However, as mentioned above, there is variability in the scaled energy of SSEs, and the data shown here cannot confirm nor deny the self-similarity of SSEs.
Some may categorize the unexpected emergence of waves from SSE zones as another synchronous phenomenon, such as episodic tremors and slips. 

Incidentally, data of \citet{tan2020connecting} (purple and orange in Fig.~1, stacked) also show the duration-dependent moment trends lower than the self-similar scaling, slightly outside the regular earthquake variability range. 
This may lead to the same conclusion for SSEs as for VLFEs, or this may be partly due to an underestimation bias caused by estimating the SSE energy only from synchronized LFEs. 
However, it should be noted that the results of \citet{tan2020connecting} are based on fitting using the Eshelby crack, which is essentially the same as estimating the prefactors of self-similar scaling if $M_0\propto t^3$ is met. 
Thus, it would be rather reasonable to view this discrepancy as representing digit-level variations in the unstacked data of \citet{tan2020connecting}.

\section{Concluding Remarks}
Guided by the Sato-Hirasawa-Matsu'ura relation (eq.~\ref{eq:SatoHirasawaMatsuura}), derived from Matsu'ura's (2024b) solution for the Sato-Hirasawa model of self-similar circular cracks, 
we have estimated the scaled energy and source parameters of various slow earthquake activities under the self-similarity hypothesis, in order to examine how plausible this hypothesis is.
Behind the Sato-Hirasawa-Matsu'ura relation is a macromechanical determinism independent of the details of faulting processes, based on the self-similarity of rupture growth. 
Hence, clearly different waveforms of regular and slow earthquakes, supposedly due to diverse fault  microphysical processes, have been described by exactly the same semi-mechanical model, the Sato-Hirasawa model~\citep{michel2019similar,supino2020self,tan2020connecting}. 
We advanced this analysis and examined the self-similarity itself from observed data and Matsu'ura's (2024b) solution (eq.~\ref{eq:Matsuura}).

One highlight of this study will be discussions on VLFEs and tectonic tremors. Their scaled energy can be well constrained from observations, and by using it via Matsu'ura's solution (eq.~\ref{eq:Matsuura}), 
we could derive the scaled energy $\Delta\tau v_{\rm r}^3$ normally difficult to constrain. 
This result paradoxically suggested that the self-similarity hypothesis may be violated for VLFEs. 
For LFEs, using eq.~(\ref{eq:Matsuura}) and the $\Delta\tau v_{\rm r}^3$ estimate obtained assuming the Sato-Hirasawa model~\citep{supino2020self}, 
we found that the scaled energy of LFEs roughly matches the observed scaled energy of VLFEs, implying that the self-similarity holds for LFEs. 
Meanwhile, 
although we estimated scaled energy from previously reported $\Delta \tau$ and $v_{\rm r}$ for self-similar SSEs, their examples are still scarce. Both $\Delta\tau$ and $v_{\rm r}$ are difficult to estimate, but SSE catalogs listing fault sizes, duration values, and moments~\citep[e.g.,][]{okada2022development} may be useful to further discuss the self-similarity of SSEs.

There may be a criticism that the theoretical model contains oversimplification. 
As mentioned in the second section, the Sato-Hirasawa model is a good starting point to consider self-similar rupture growth, but there are exceptions.
Theoretical models of \citet{sato1973body}, \citet{kanamori1975theoretical}, and \citet{kanamori1977energy} 
have consistently simplified the analyses by approximating elastodynamic slip distributions at subshear speeds with Eshelby's (1957) elastostatic solution~\citep{kaneko2015variability}. The effect of Mach waves emitted from the super-shear rupture with singular directivity~\citep{dunham2008attenuation} may be an exception.
The rupture arrest process evidently violates the self-similarity and is therefore not captured: the artificial stopping phase in the Sato-Hirasawa model~\citep{madariaga1976dynamics} and similar flaws will become an issue when improving the present discussion. 


Once again, one key finding of this study is that the self-similarity hypothesis can be tested by comparing the scaled energy and moment-duration plots from the perspective of source mechanics. 
For VLFEs, the duration is about 1--2 digits larger than the prediction of self-similarity, and the self-similarity hypothesis is at least apparently violated. 
We need to be cautious about the fact that $M_0\propto t^3$ holds if self-similar, but the reverse is not true. 
In order for the self-similarity interpretation to survive, the self-similarity violation of VLFEs probably has to be seen as measurement bias in a broad sense. Only careful data analysis can settle whether it is truly self-similar as it is also quite possible that estimation uncertainty is the reason why the VLFEs violate the theoretical consequence of self-similarity (the Sato-Hirasawa-Matsu'ura relation, eq.~\ref{eq:SatoHirasawaMatsuura}). 
Our scaled energy estimation of LFEs, consistent with the observed scaled energy of VLFEs, also needs to be tested from direct measurements of the scaled energy of LFEs. 
Certainly, it is natural to regard $M_0\propto t^3$ as a sign of self-similarity, and we ourselves did so when estimating the source parameters and scaled energy, but conversely, as this study did, 
we can test this hypothesis itself by examining those results with caveats, thanks to Matsu'ura's (2024b) solution.

It was a surprise to us that another cluster of \citet{michel2019similar} could be included in one of those two modes found by \citet{tan2020connecting}. 
Coincidence of their results may also suggest that measuring SSE moments by LFEs is not that inferior to the geodetic direct measurements of SSE moments. 
However, that of \citet{maeda2009spatiotemporal} with observed scaled energy far larger than those clusters' may be the third mode, possibly refuting the self-similarity of SSEs.

Although questions about actual events thus arise, there is little theoretical doubt that there is an identical scaling for self-similar slow and fast earthquake growth to follow. 
The deduced kinematics (eq.~\ref{eq:moment_radiation_ratio_relation}) is evocative and, above all, systematizes disparate seismic and aseismic slips via the mechanics of self-similarity. 
The Sato-Hirasawa-Matsu'ura relation (\ref{eq:SatoHirasawaMatsuura}) unifying the self-similar rupture growth is useful for deeper considerations of $M_0\propto t^3$ rupture phenomena, including whether they are truly self-similar.

\section*{List of abbreviations}
LFEs: low-frequency earthquakes; VLFEs: very-low-frequency earthquakes; SSEs: slow slip events.

\section*{Declarations}

\section*{Availability of data and materials}

The data underlying this paper are available from \citet{allmann2009global}, 
\citet{ide2018seismic}, 
\citet{michel2019similar}, \citet{tan2020connecting}, and 
\citet{DVN_HCWJUI_2020}.

\section*{Competing interests}
The author has no competing interests.

\section*{Funding}
This study was supported by JSPS KAKENHI Grant Number 23K19082. 

\section*{Authors' contributions}
This is a single-authored article. 

\acknowledgments{
The author deeply appreciates the courtesy of Dr. Mitsuhiro Matsu'ura, who let the author use the unpublished result of \citet{matsuura2024A} and notice the widely spread \citet{haskell1964total} convention. The author also acknowledges comments from Dr. Tomoaki Nishikawa that greatly improved the description of slow earthquakes in this article. 
He also thanks Dr. Takane Hori and Dr. Yukitoshi Fukahata for their encouraging comments. 
}

\begin{figure*}
   \includegraphics[width=150mm]{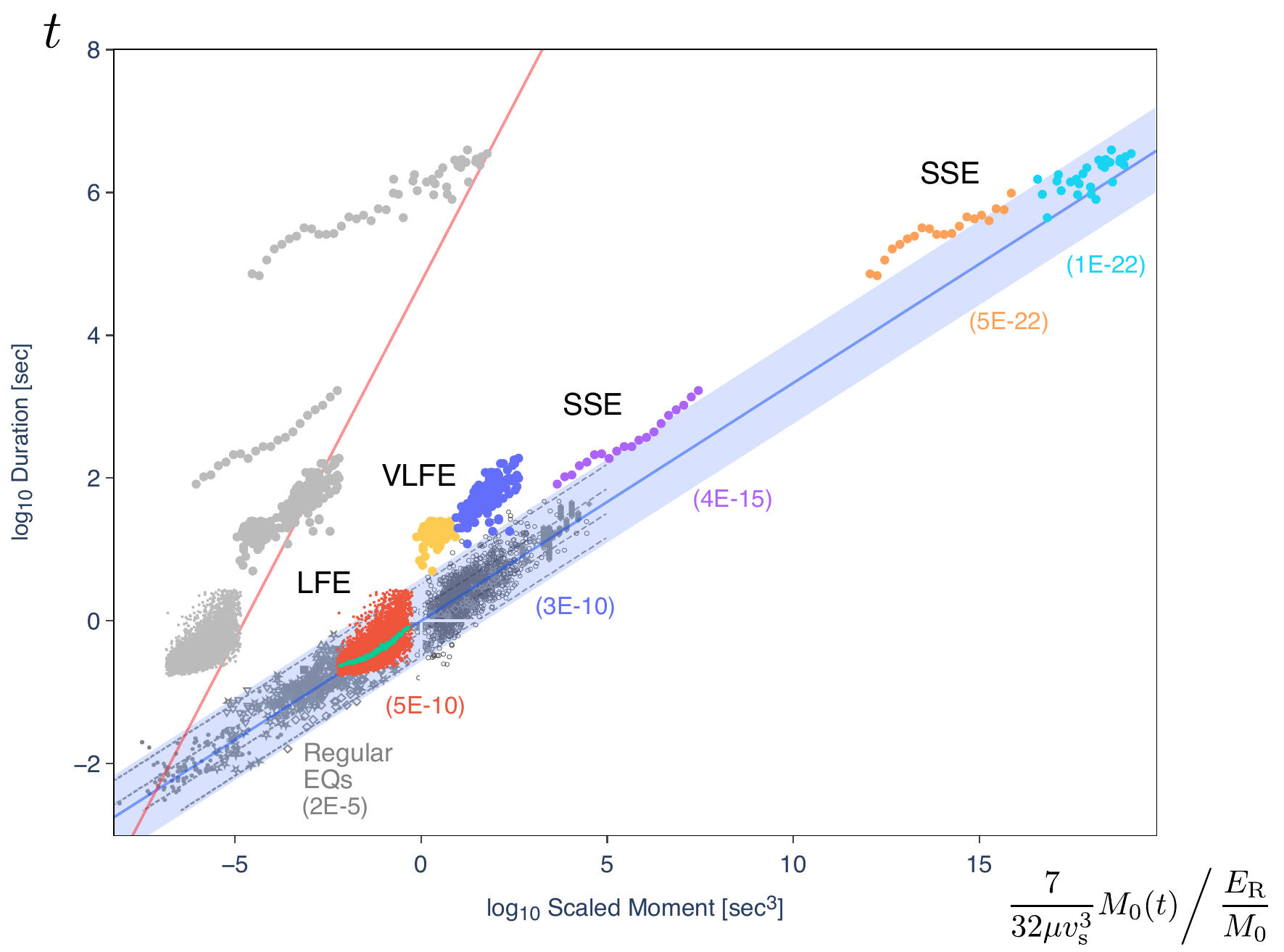}
  \caption{
A moment-duration plot of various regular and slow earthquakes rescaled for self-similar rupture growth. 
The horizontal axis takes the scaled moment $[7M_{\rm 0}/(32\mu v_{\rm s}^3)]/(E_{\rm R}/M_0)$.
The vertical axis takes duration $t$. 
The numbers in parentheses represent nominal estimates of the scaled energy for respective clusters. 
For slow earthquake members other than VLFEs, the scaled energy is estimated by assuming self-similarity. 
For VLFEs, moments are rescaled by using the observed scaled energy to see if they satisfy self-similarity scaling. See text for details.
The rescaling assumes $\mu=30$ GPa and $v_{\rm S}=4$ km/sec.
Regular earthquake data are compiled by \citet{allmann2009global}. 
VLFE data are distributed by \citet{ide2018seismic} (stacked: yellow; unstacked: blue). 
LFE data are downloaded from \citet{DVN_HCWJUI_2020} and then converted from frequency to duration, assuming the conversion relation used by \citet{supino2020self} (stacked: green; unstacked: red). 
SSE data are manually read from \citet[cyan, unstacked]{michel2019similar} and \citet[purple and orange, stacked]{tan2020connecting}.
For comparison, the relative positions of slow earthquakes from regular earthquakes, before rescaling, are shown in grey, together with the scaling of self-similar rupture growth (blue) and the diffusive scaling bounding the upper limit of the final size of slow earthquakes~\citep[pink]{ide2007scaling}. 
The bandwidth for self-similar rupture scaling is taken at about $\pm10^{1.5}$ [N$\cdot$m] as that for regular earthquakes by \citet{allmann2009global}. 
}
\end{figure*}

\end{document}